%% file: enhanced_excitation_190805.tex
\documentclass[%
reprint,
superscriptaddress,
 amsmath,amssymb,
 aps,
 pra,
]{revtex4-1}
 
\usepackage{amsmath}
\usepackage{amssymb}
\usepackage{amsfonts}
\usepackage{dsfont}
\usepackage{graphicx}
\usepackage[utf8]{inputenc}
\usepackage[T1, T2A]{fontenc}
\usepackage[english]{babel}
\usepackage{xcolor}

\newcommand{\p}{\partial}
\newcommand{\avg}[1]{\left\langle #1 \right\rangle}
\newcommand{\bigexp}[1]{\exp{\left\{ #1 \right\}}}
\DeclareMathOperator{\acosh}{acosh}
\let\Re\relax
\DeclareMathOperator{\Re}{Re}
\let\Im\relax
\DeclareMathOperator{\Im}{Im}

\begin{document}
\title{Enhanced excitation of a driven bistable system induced by spectrum degeneracy}
\date{\today}
\author{Evgeny V. Anikin}
\affiliation{Skolkovo Institute of Science and Technology, 121205 Moscow, Russia}
\author{Natalya S. Maslova} 
\affiliation{
    Department of Quantum electronics and Quantum Center, 
                        Faculty of Physics, Moscow State University, 119991 Moscow, Russia
    }
\author{Nikolay A. Gippius}
\affiliation{Skolkovo Institute of Science and Technology, 121205 Moscow, Russia}
\author{Igor M. Sokolov}
\affiliation{Institut für Physik and IRIS Adlershof, Humboldt Universität 
             zu Berlin, Newtonstraße 15, 12489 Berlin, Germany}
\begin{abstract}
    The non--equilibrium statistics and kinetics of a simple bistable system
    (resonantly driven nonlinear oscillator coupled to reservoir) have been investigated by
    means of master equation for the density matrix and quasiclassical Fokker--Planck equation
    in quasienergy space. We found out that the system's statistical and kinetic properties 
    drastically change when the quasienergy states become nearly degenerate and the occupation
    of the most excited state is strongly enhanced.
    It has been revealed 
    that in nearly degenerate case a new critical quasienergy parameter emerges. Below the critical 
    quasienergy value the eigenstates are superpositions of the quasiclassical states from different 
    phase space regions, while above this value the eigenstates correspond to only one particular
    region of the phase space. We have also generalized Keldysh theory for ionization of atoms
    in the electromagnetic field for bistable systems. It has been demonstrated that Keldysh parameter
    in bistability region is large when pumping intensity is smaller than the critical value. It has
    been shown by direct calculations that multi--photon transition amplitude coincides with the
    tunneling amplitude. So, multi--photon transitions and tunneling between the regions of the 
    phase space are just the same effects. We also demonstrated that for bistable systems the Keldysh
    parameter logarithmically depends on the external field amplitude.
\end{abstract}
\maketitle
\section{Introduction}
Nowadays bistability is one of the most pronounced phenomena in modern optics and electronics.
It has broad applications in all--optical logic and memories performance. 
So, controllable changing
of different stable states occupation and the control of transition rates between them are among
the most important problems. 
The solution of these problems 
relies on one's knowledge of optimal perturbation which transfers the system
from one stable state to another one. Another problem is how to control the structure of stable 
states by changing the system parameters such as external field frequency and intensity.
Bistability has been widely studied in different experimental setups: cold clouds of atoms inside the 
optical cavity \cite{Rosenberger1991}, \cite{Wang2001}, 
exciton--polariton modes in microcavities with external pumping \cite{Gippius2007}, 
fiber ring cavities \cite{Li2017} and 
mesoscopic Josephson junction array resonators \cite{Muppalla2018} in the external field, etc.
The results of many of these experiments can be understood by investigation of bistable single--mode
system with Kerr--like nonlinearity.
Resonance response of a bistable single--mode system with a Kerr--like 
nonlinearity to the external field is described by a 
model of a driven nonlinear oscillator interacting with dissipative environment \cite{Drummond1980}. 
This model also 
describes an atomic system with several energy levels coupled to the cavity mode after 
adiabatically excluding atomic variables \cite{Gothe2019}, \cite{Shirai2018}, \cite{Dong2019}.

Moreover, the model of a driven nonlinear oscillator 
is a minimal model of a bistable driven system out of equilibrium.
In the quasiclassical limit, its statistics and kinetics can be 
described by a Fokker--Planck equation (FPE) \cite{Maslova2007}. By means of FPE, it is possible to
find the stationary distributions, relaxation rates and the occupations of two classical stable
states of the oscillator. For the quantum oscillator, the 
non--equilibrium statistics and relaxation kinetics at different temperatures 
were studied numerically using the rate equation \cite{Risken1987}, \cite{Risken1988}. 
Also for the quantum oscillator 
the exact Glauber--Sudarshan function of the steady state 
was obtained \cite{Drummond1980} in the case of zero bath temperature. 


In the described model of the bistable driven system, 
it has been shown \cite{Maslova2007} that there exist different regions of the 
classical phase space with degenerate energies. 
Among the quantum effects, the effects of tunneling transitions between these regions are of
particular interest. As mentioned in \cite{Maslova2019}, tunneling increases the population of the 
higher amplitude state, which is also a squeezed state. Also it increases the relaxation rate to 
the stationary distribution. From the quasiclassical point of view, tunneling can lead to 
hybridization of quasiclassical states from different regions of the classical phase plane. It can
be shown to be very strong in the case of integer or half--integer 
detuning--nonlinearity ratio, when the quasiclassical states from the different regions of the phase
space become degenerate. This also corresponds to the multi--photon resonance between the 
real energy levels of the driven system.
The hybridization of the quasienergy states from different regions of the phase space in the case
of multi--photon resonance can strongly change the kinetics of the considered system. 
However, the kinetics of the considered bistable system 
with the eigenstates which are superpositions of the states from different regions
of the classical phase space drastically differs from the non--degenerate case investigated previously
\cite{Maslova2019} and are not studied yet.

In the case of strong hybridization between the states from different regions of the phase plane,
the transition rate between them can be explained by generalization of Keldysh theory for
ionization of atoms in electromagnetic field \cite{Keldysh1965} for bistable systems. 
The Keldysh theory explains the interplay between tunneling and multi--photon ionization. 
However, for bistable systems the correspondence between multi--photon transitions 
and tunneling effects is not clear. To understand it, 
one should define the Keldysh parameter $\gamma_K$ as 
the ratio of <<tunneling time>> to the period of motion along the classical trajectory.
It will be demonstrated that when the field intensity is much smaller than the critical
value defining the range of bistability, the Keldysh parameter is large,
$\gamma_K \gg 1$ and tunneling probability is just the same as 
the probability of multi--photon excitation. It will be also shown 
that $\gamma_K$ logarithmically depends on
the strength of the external field $f$ while in the case of ionization of atoms it is 
proportional to $f^{-1}$.

\section{The simple model of bistable driven system}
The Hamiltonian of a driven resonant mode with Kerr--like nonlinearity
in the rotating--frame approximation reads
\begin{equation}
    \label{quant_bist_ham}
        \hat{H} =-\Delta \hat{a}^\dagger \hat{a} + 
        \frac{\alpha}{2}(\hat{a}^\dagger \hat{a})^2 + f(\hat{a} + \hat{a}^\dagger).
\end{equation}
Here $\Delta$ is the detuning between the driving force and the resonant
oscillator frequency, $\alpha$ is the blue shift due to nonlinearity, and $f$ is proportional
to the amplitude of the driving force.

\begin{figure}[h]
    \includegraphics[width=\linewidth]{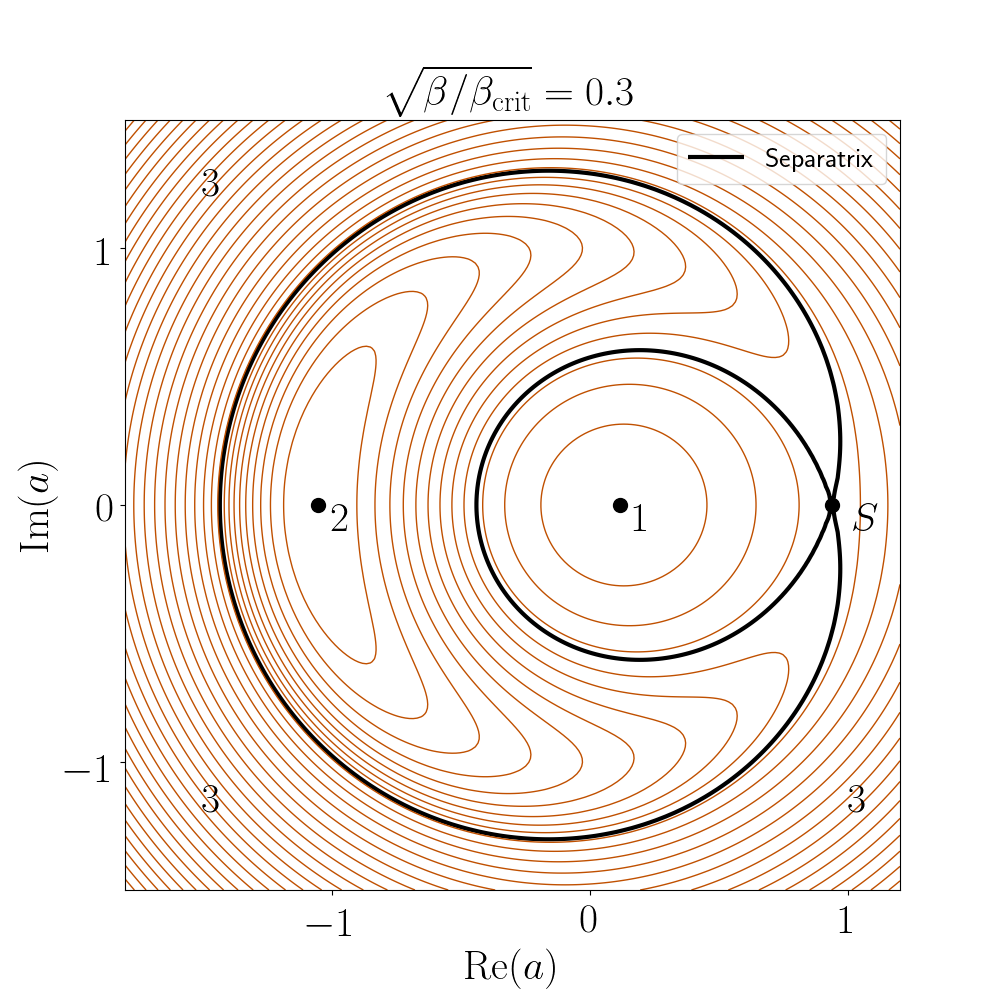}
    \caption{The phase portrait of the nonlinear oscillator with 
             Hamiltonian \eqref{quant_bist_ham} where the operators $\hat{a}, \hat{a}^\dagger$ 
             are replaced by c--numbers. The parameters are $\Delta = \alpha = 1$, 
             $\sqrt{\beta/\beta_\mathrm{crit}} = 0.3$. The stationary points 1, 2 and S are
             shown by black dots. The separatrix is denoted by a thick black line. It divides 
             the phase plane into regions which are shown by numbers 1, 2, 3.
             }
    \label{fig:phase_portrait}
\end{figure}

In the classical limit, one should replace the operators $\hat{a}$, $\hat{a}^\dagger$ 
in \eqref{quant_bist_ham} with classical field amplitudes 
$a, a^*$ to obtain the classical Hamiltonian.  
The classical phase portrait of the system is shown on Fig.~\ref{fig:phase_portrait}: the classical
trajectories in the $a$ plane are given by the contour lines of the classical Hamiltonian. Each
classical trajectory corresponds to a certain quasienergy value $\epsilon$.
The only dimensionless parameter governing the system classical dynamics 
is $\alpha f^2/\Delta^3 \equiv \beta$,
\textcolor{black} {
which can be treated as the rephasing rate of 
the nonlinear driven oscillator \cite{Lamb1972}. This parameter can also be identified with
the Dicke cooperation parameter determining the typical rate of the intensity growth of a
superradiance pulse. }
Bistability appears when $0 < \beta < 4/27$.

For the quantum Hamiltonian, there exists another dimensionless parameter 
$m \equiv 2\Delta/\alpha$. The quasiclassical limit is acquired at large $m$.

A key feature of the driven nonlinear oscillator is the presence of two stable stationary states
which means bistability. Another important feature is the presence of 
a self--intersecting trajectory called separatrix, which divides the phase plane
into three regions $1$, $2$ and $3$ and passes through the unstable stationary point $S$.
The regions $1$ and $2$ contain the stationary states with 
smaller and larger amplitude respectively. The quasienergies of the states 1,2,S are denoted 
by $\epsilon_1$, $\epsilon_2$ and $\epsilon_\mathrm{sep}$, and they always obey the 
inequality $\epsilon_2 < \epsilon_\mathrm{sep} < \epsilon_1$. In further discussion, we will also 
use the dimensionless quasienergy defined as $E = \alpha\epsilon/\Delta^2$, 
$E_1 = \alpha\epsilon_1/\Delta^2$,
$E_2 = \alpha\epsilon_2/\Delta^2$,
$E_\mathrm{sep} = \alpha\epsilon_\mathrm{sep}/\Delta^2$,


The states of the quantum Hamiltonian in the limit of 
large numbers of excitation quanta  
correspond to a discrete set of classical trajectories on the phase portrait. The
corresponding energies can be obtained from the Bohr--Sommerfeld quantization rule.

The interaction with the environment can be described by the following interaction Hamiltonian:
\begin{equation}
    H_\mathrm{int} = \hat{\xi}^\dagger \hat{a} + \hat{\xi} \hat{a}^\dagger,
\end{equation}
where $\hat{\xi}, \hat{\xi}^\dagger$ are the operators of random force with correlation functions 
defined as
\begin{equation}
    \begin{gathered}
        \avg{\hat{\xi}(t)\hat{\xi}^\dagger(t')} = \gamma(N + 1)\delta(t-t'),\\
        \avg{\hat{\xi}^\dagger(t)\hat{\xi}(t')} = \gamma N\delta(t-t'),\\
    \end{gathered}
\end{equation}
where $N$ is the number of noise quanta and $\gamma$ is the damping caused by the interaction with 
the environment.
The kinetics of the quantum bistable driven oscillator in the limit of weak coupling with the
environment can be treated in the diagonal approximation for the density matrix. 
In this approximation,
one obtains the rate equation dealing with probabilities $P_n$ of occupation of 
the $n$--th quasienergy state:
\begin{equation}
    \label{rate_eq}
    \begin{gathered}
        \frac{dP_n}{dt} = \sum_{n'} w_{nn'} P_{n'} - w_{n'n} P_n,\\
        w_{nn'} = \gamma
                \left[ 
            (N + 1)| \langle n | \hat{a} | n' \rangle|^2 
           + N |\langle n' | \hat{a}| n \rangle |^2 \right].
    \end{gathered}
\end{equation}
If each quasienergy state can be uniquely attributed to one of 
the regions of the phase space,
in the limit of large number of excitation quanta
the rate equation transforms to the classical Fokker--Planck equation
in the quasienergy space: 
\begin{equation}
    \label{energy_fokker_planck}
    \begin{gathered}
        \frac{\p P_i(E)}{\p t} = \frac{1}{T_i(E)} \frac{\p J_i(E)}{\p E}, \\
        J_i(E) = \vartheta K_i(E) P_i(E) + 
            Q D_i(E) \frac{\p P_i}{\p E}.
    \end{gathered}
\end{equation}
where $T_i(E)$ is the period of motion along the classical trajectory, the coefficients 
$K_i(E)$ and $D_i(E)$ are the drift and diffusion coefficients in quasienergy space. 
The coefficients $T_i(E), K_i(E)$ and $D_i(E)$ are defined in the 
Appendix~\ref{appendix:quasiclassical_quantization}. The dimensionless parameters $\vartheta, Q$ are
defined as
$\vartheta = \gamma/\Delta$ and $Q = \vartheta\alpha(N + 1/2)/\Delta$.
The probability densities $P_i(E)$ are the continuous limits of $P_n$, where the probabilities 
$P_n$ are considered as functions of dimensionless quasienergy $E \equiv \alpha\epsilon_n/\Delta^2$,
and $i$ denotes the region of the phase space according to Fig.~\ref{fig:phase_portrait}. The
function $P_2(E)$ is defined for $E_2 < E < E_\mathrm{sep}$, $P_1(E)$ for 
$E_\mathrm{sep} < E < E_1$, and $P_3(E)$ for $E > E_\mathrm{sep}$.

However, as the quasienergy states in regions 1 and 3 can have same energies,
it is possible that the true quasienergy states are superpositions of the quasiclassical states 
from regions 1 and 3 due to quantum tunneling. So, the classical Fokker--Planck equation should be
generalized to take this effect into account.
It will be shown that hybridization of quasiclassical
states from regions 1 and 3 strongly modifies the non--equilibrium statistics and kinetics of the
system.
\section{The structure of quasienergy states}
\label{sec:states_structure}
In this section, we will consider the model quasi--classically, although the results obtained here are
valid beyond the quasiclassical approximation. Within the quasiclassical approach, the eigenstates of
the quantum Hamiltonian correspond to a discrete set of trajectories which are obtained using the 
Bohr-Sommerfeld rule:
\begin{equation}
    \label{ad_inv}
    \frac{1}{2\pi}\oint p\,dq = 2\pi n.
\end{equation}
The variables $q,p$ are the canonical coordinate and momentum defined 
by $a,a^* = \frac{q \pm ip}{\sqrt{2}}$. For the states from region 2, the value of quasienergy uniquely
defines the classical trajectory. However, if a state has quasienergy 
$\epsilon_\mathrm{sep} < \epsilon < \epsilon_1$, it can lie either in the region 1 or region 3. 
Therefore if the quasienergies of some states obtained from the Bohr--Sommerfeld rule 
are close enough, the true eigenstates of the quantum Hamiltonian 
can be superpositions of the quasiclassical states due to quantum tunneling.
\begin{figure}[h]
    \includegraphics[width=\linewidth]{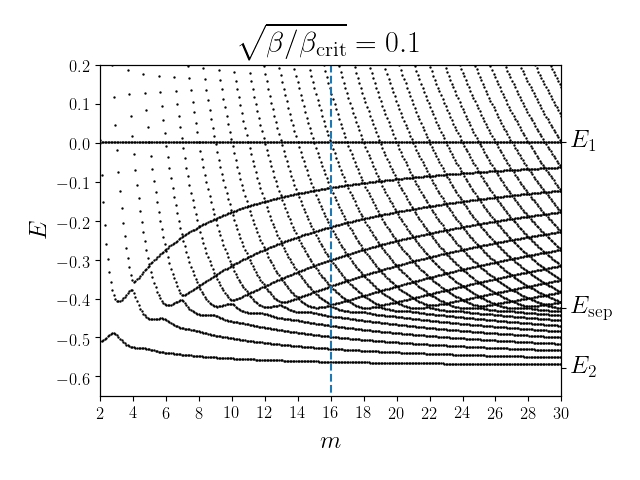}
    \caption{The dependency of the quantum driven nonlinear oscillator quasienergy levels on
             $m$. Different behavior of levels above and below $E_\mathrm{sep}$ is 
             evident: for $E_\mathrm{sep} < E < E_1$, there are two families of 
             almost intersecting (anti--crossing)
             lines which correspond to the quasiclassical states from regions $1$ and $3$. 
             All anti--crossings occur at integer 
             values of $m$: the blue dashed line corresponds to $m = 16$.
             For $E_2 < E < E_\mathrm{sep}$, there is only one family of lines which corresponds to
             the quasiclassical states from region $2$.}
    \label{fig:exact_diag}
\end{figure}

Let the set of quasiclassical states from regions 1 and 3 be $|n_{1}\rangle$, $|n_{3}\rangle$ with
corresponding quasienergies $\epsilon_{n_{1}}$, $\epsilon_{n_3}$. If for some
$n$, $n'$ the quasienergies are almost equal, $\epsilon_{n_1} = \epsilon_{n'_3}$, the quasiclassical
states $|n_1\rangle$ and $|n'_3\rangle$ form superpositions. Formally, for each $n,n'$
it is possible to find the parameters $\beta$, $m$ such as $\epsilon_{n_1} = \epsilon_{n'_3}$. 
In principle, these values of $\beta$ and $m$ could be different for different pairs $n, n'$. 
However, a special feature of the Hamiltonian \eqref{quant_bist_ham} is that the degeneracy of the
quasiclassical states occurs exactly at integer values of $m$ for all $\beta$ 
in the region of bistability.
Therefore at each integer value of $m$ all quasiclassical states from regions 1 and 3 can
be grouped in pairs so that within each pair the values of quasienergy are equal. When $m$ is
close to an integer, the states still can be grouped in pairs with close values of quasienergies.
This is clearly seen from the exact diagonalization of the Hamiltonian 
\eqref{quant_bist_ham}. On the Fig.~\ref{fig:exact_diag}, the eigenvalues of
the Hamiltonian \eqref{quant_bist_ham} are shown as functions of $m$ at constant $\beta$. It can
be seen that for $E_\mathrm{sep} < E < E_1$, there exist two families of lines which correspond 
to the states from regions 1 and 3. The anti--crossings of these lines indicate the degeneracy of
the quasiclassical states. It is evident that all these anti--crossings occur exactly at integer values
of $m$.

To put the statements of the previous paragraph on the theoretical ground, let us first consider 
the case $f = 0$. At $f = 0$, the Hamiltonian commutes with 
the number of excitation quanta operator $a^\dagger a$, and eigenstates have the form 
$|n\rangle$ with corresponding quasienergies
\begin{equation}
    \label{unperturbed_levels}
    \begin{gathered}
        \epsilon_n^{(0)} = -\Delta n + \frac{\alpha}{2}n^2 = 
        \frac{2\Delta^2}{\alpha}\frac{n}{m}\left(\frac{n}{m} - 1\right)\\
        m \equiv \frac{2\Delta}{\alpha}.\\
    \end{gathered}
\end{equation}
At integer $m$, it is clear that all quasienergy levels split into pairs with same energy, as
$\epsilon_n^{(0)} = \epsilon_{m-n}^{(0)}$.  

Let us proceed to the case $f > 0$.  Let us consider the pair of states $|n\rangle$ and 
$|m-n\rangle$, $n < m/2$ which are degenerate at $f=0$. From the 
quasiclassical point of view, these states corresponds to circular trajectories on the phase plane 
with different radii. From the Bohr--Sommerfeld rule, it follows that the values of adiabatic invariant
$\frac{1}{2\pi}\oint pdq$ for these states are $-n$ and $m-n$. Then let us switch on the external
field adiabatically. After that, the quasienergies of the states change, 
but the values of the adiabatic invariant remain the same. The corresponding trajectories lie 
in regions 1 and 3 respectively unless $f$ is so large that they merge into a single trajectory
from region 2.
Therefore the quasienergies of the resulting states are $\epsilon_1(-n, \beta)$ 
and $\epsilon_3(m-n, \beta)$, where the energies are understood as 
functions of the adiabatic invariant.

For the considered model, it is possible to prove the identity
\begin{equation}
    \epsilon_1(-n, f) = \epsilon_3(m-n, f),
\end{equation}
or equivalently
\begin{equation}
    \label{ad_inv_identity}
    n_3(\epsilon, f) - n_1(\epsilon, f) = m.
\end{equation}
This happens because $n_1(E, f)$ and $n_3(E, f)$ have 
an analytic expression through the same elliptic integral with different contours
(see Appendix \ref{appendix:quasiclassical_quantization}). With the same reasoning, it is also
easy to verify that the classical periods of the trajectories from regions 1 and 3 with equal 
quasienergies are the same \cite{Dykman2005}.
Thus, in the quasiclassical limit the energies of the states with numbers $n$ and 
$m-n$ obtained from Bohr--Sommerfeld rule remain the same 
even at finite $f$. This supports the statement that the degeneracy of the quasiclassical states
from regions 1 and 3 happens simultaneously for all states at integer values of $m$.

To find the eigenstates in the case when the quasiclassical states from regions 1 and 3 are
degenerate, it is necessary to find the tunneling amplitude,
which can also be obtained quasiclassically. 
For that, one should consider the motion of the system in the classically forbidden
area. The quasiclassical tunneling exponent equals half of the action of a 
classical closed trajectory in 
imaginary time. The resulting amplitude is
\begin{equation}
    \label{quasiclassical_splitting}
        \omega_R(E) \sim \frac{1}{T(E)} e^{-S_\mathrm{tunn}(E)},\\
\end{equation}
\begin{multline}
    \label{tunneling_exp}
    S_\mathrm{tunn}(E) = \frac{1}{2}\oint p_\mathrm{im}dq =\\
         = \frac{m}{2}\int_{s_1}^{s_2} 
             \acosh\left\{ \frac{E + \frac{s^2}{2} - \frac{s^4}{8}}
                        {\sqrt{2\beta}s}\right\} s\,ds,
\end{multline}
where $p_\mathrm{im}$ is defined from the equation $H(q, ip_\mathrm{im}) = \epsilon$. When 
$\beta \ll \beta_\mathrm{crit}$, 
\begin{equation}
    \label{tunneling_exp_1}
    S_\mathrm{tunn}(E) = \frac{m}{2}\ln{\frac{1}{\beta}}\sqrt{1 + 2E} + O(1).
\end{equation}
For nearly degenerate states, when $E_n \approx E_{m-n}$, the tunneling exponent takes the form
\begin{equation}
    \label{tunneling_exp_2}
    S_\mathrm{tunn}(E_n) = \frac{m}{2}\ln{\frac{1}{\beta}}\left(1 - \frac{2n}{m}\right),
\end{equation}
which is especially convenient for comparison with multi--photon transition amplitude obtained by
perturbation theory.

\begin{figure}[h]
    \centering
    \includegraphics[width=\linewidth]{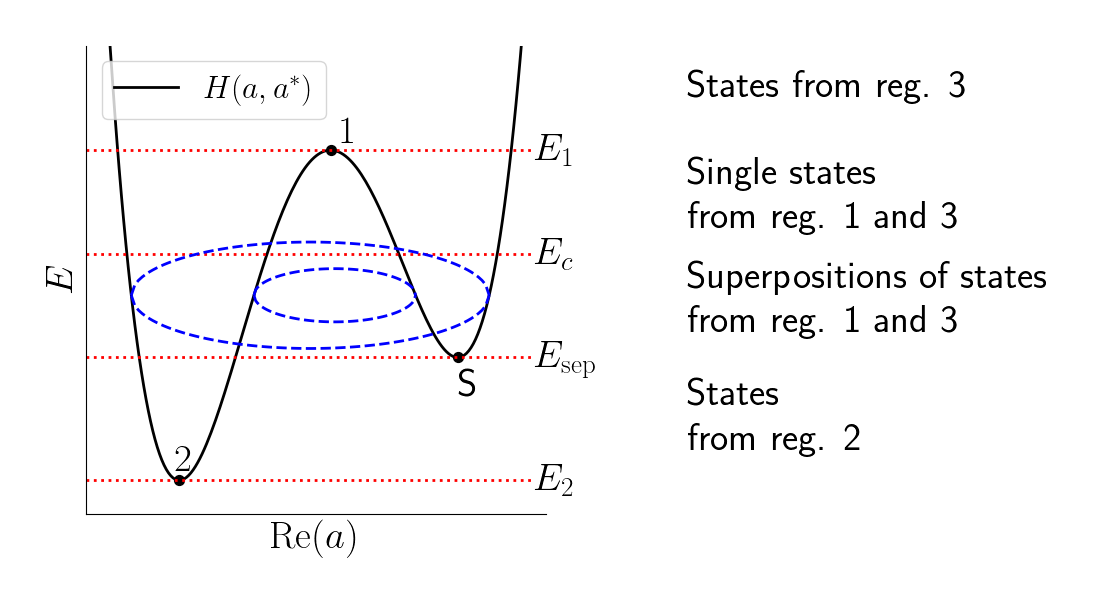}
    \caption{The quasienergy ranges containing quasienergy states with different structure
             are shown.
             The black thick line denotes the Hamiltonian function $H(a, a^*)$ at
             $\Im{a} = 0$. Its extrema correspond to the stationary states 1, 2 and S. Also
             a quantum state which is a superposition of two quasiclassical states is
             schematically shown by a blue dashed line.
             }
    \label{fig:states_structure}
\end{figure}

The eigenstates with account for tunneling between the nearly--degenerate eigenstates 
can be found from an effective 
two--level Hamiltonian for two near--degenerate quasiclassical states. It has form
\begin{equation}
    \label{eff_ham_13}
    H_n = 
    \begin{pmatrix}
        \epsilon_{n_1} & -\omega^n_R\\
        -\omega_R^n & \epsilon_{n_3}
    \end{pmatrix},
\end{equation}
where $\epsilon_{n,1}$ and $\epsilon_{n,3}$ are the quasienergies 
of the quasiclassical states $|n_1\rangle$, $|n_3\rangle$, 
and $\omega^{n}_R$ is the tunneling amplitude.
When $m$ slightly deviates from an integer, $m = m_0 + \delta m$,
the difference of quasiclassical energies is
\begin{equation}
    \label{energy_diff}
    \delta \epsilon_n = \epsilon_{n_1} - \epsilon_{n_3} = -\frac{2\pi \delta m}{T(E_n)}.
\end{equation}
The eigenstates of the effective Hamiltonian are
\begin{equation}
    \label{pm_states}
    \begin{gathered}
       |n_+\rangle = c_n|n_1\rangle + s_n|n_3\rangle,\\
       |n_-\rangle = -s_n|n_1\rangle + c_n|n_3\rangle.\\
    \end{gathered}
\end{equation}
where the coefficients $c_n, s_n$ are defined by the ratio between $\omega_R$ and 
$\delta \epsilon_n$:
\begin{equation}
    \begin{gathered}
        c_n \equiv \cos{\theta_n},\quad s_n \equiv \sin{\theta_n},\\ 
        \tan{2\theta_n} = \frac{2\omega_R}{\delta \epsilon_n} = 
            \frac{1}{\pi \delta m} e^{-S_\mathrm{tunn}(E_n)}.
    \end{gathered}
\end{equation}
Because of exponential dependence of $\omega_R$ on $E$, the coefficients
$c_n$ and $s_n$ have a step--like dependence on $n$. When
$\omega_R(E_n) \ll \delta \epsilon_n$, 
$s_n \approx 0,\quad c_n \approx 1$, and the eigenstates correspond to 
distinct trajectories in the regions 1 and 3 of the phase portrait. On the contrary,
when $\omega_R(E_n) \gg \delta \epsilon_n$, $s_n \approx c_n \approx 1/\sqrt{2}$, 
and the states are very close to symmetric and antisymmetric superpositions of trajectories. 
These alternatives are separated by the critical value of quasienergy $E_c$ for which
$\omega_R(E_c) \approx \delta \epsilon(E_c)$. 
Combining this with Eq.~\eqref{energy_diff}, one
get the equation for $E_c$
\begin{equation}
    \label{E_c_equation}
    \frac{1}{\pi \delta m} e^{-S_\mathrm{tunn}(E_c)} = 1.
\end{equation}
The resulting structure of quasienergy states is schematically depicted on 
Fig.~\ref{fig:states_structure}. Below $E_\mathrm{sep}$, there are only
states from classical region 2. Between $E_\mathrm{sep}$ and $E_c$, 
the states from regions 1 and 3 form superpositions,
and above $E_c$, the states from regions 1 and 3 don't hybridize.

\section{The symmetry of the Hamiltonian}
\label{sec:symmetry}
The quasiclassical arguments of the previous section are valid only in the leading order in the 
quasiclassicity parameter $m$. In particular, the quasienergy values of the system can be expanded 
in asymptotic series in $1/m$, and the Bohr--Sommerfeld rule gives only the leading term of these
series. Therefore additional arguments are needed to explain the simultaneous anti--crossings 
of the quasienergy levels at integer $m$. 

The rigorous proof can be given using the perturbation theory in $f$ for the quantum Hamiltonian
\eqref{quant_bist_ham}. At $f=0$, the Hamiltonian commutes with $\hat{n} = a^\dagger a$,
so the quasienergies of the Hamiltonian \eqref{quant_bist_ham} are given by 
\eqref{unperturbed_levels}. At small $f$, one can use the perturbation theory 
to find the corrections to the energy of the state $|n\rangle$. 
For example, the second--order correction is
\begin{equation}
    \label{2_ord_corr}
     \delta \epsilon^{(2)}_n = \frac{f^2}{\alpha}\cdot  \frac{(m + 1)}{(m - 2n)^2  - 1}.
\end{equation}
The perturbative correction $\epsilon^{(2)}_n$ 
is symmetric with respect to replacement $n \to m - n$. 
This is in agreement with the quasiclassical arguments of Sec.~\ref{sec:states_structure}: it was 
shown that the changes of quasiclassical quasienergy due to adiabatic change of 
$f$ are the same for states $|n\rangle$ and $|m-n\rangle$. This statement is not 
only valid in the quasiclassical framework but also holds exactly in the second order of 
perturbation theory. Moreover, we obtained a rigorous proof 
that the same holds for higher orders $k$ of perturbation theory, 
$\epsilon_n^{(k)} = \epsilon_{m-n}^{(k)}$
up to order $m - 2n$. The complete proof is given in the Appendix~\ref{appendix:symmetry}.
For integer $m$ one should utilize the degenerate perturbation theory which also 
takes into account multi--photon transitions between the states $|n\rangle$ and $|m-n\rangle$. 
These transitions occur only in order $m - 2n$, and up to this order, the non--degenerate
perturbation theory remains valid. Therefore the quasienergy splitting between the 
states $|n\rangle$ and $|m-n\rangle$ occurs only in the order $m - 2n$.

\section{The generalization of Keldysh theory 
         for ionization of atoms in 
         electromagnetic field}
Splitting between the quasienergy states and the transitions between the 
regions of the phase space can be treated not only as tunneling between the regions of the phase 
space but also in $(m - 2n)$--th order of perturbation theory in $f$, as mentioned in 
Sec.~\ref{sec:symmetry}. Both approaches lead to just the same effects. Such behavior can be 
understood in the frame of Keldysh theory for ionization of atoms in electromagnetic field
generalized for bistable systems with discrete spectrum.
Transition amplitude depends on Keldysh parameter $\gamma_K \equiv T_\mathrm{im}(E)/T(E)$ where 
$T_\mathrm{im}$ is the <<tunneling time>> which is defined as
$T_\mathrm{im} \equiv (\alpha/\Delta^2)\p_ES_\mathrm{tunn}$, where 
$S_\mathrm{tunn}(E)$ is the tunneling action defined by \eqref{tunneling_exp_1}.
The <<tunneling time>> has the meaning of the half--period of motion along the closed trajectory
in the imaginary time.  From the expression \eqref{tunneling_exp_2}, we obtain that 
$T_\mathrm{im}(E) \sim \Delta^{-1}\ln{\frac{1}{\beta}}$. The period in the real time 
is always $\sim \Delta^{-1}$ unless the quasienergy is close to $E_\mathrm{sep}$. 
So, $\gamma_K \sim \ln{\frac{1}{\beta}} \gg 1$ for $\beta \ll \beta_\mathrm{crit}$. In this limit,
the tunneling amplitude coincides with $(m-2n)$--order perturbation theory
multi--photon transition amplitude.
Indeed, as it was shown in \cite{Maslova2019}, the multi--photon transition amplitude reads
\begin{multline}
    \label{pert_splitting}
    \omega_R^{n,m-n} = \frac{V_{n,n+1}\dots V_{m-n-1,m-n}}
                            {(\epsilon_n^{(0)} - \epsilon_{n-1}^{(0)})
                            \dots(\epsilon_n^{(0)} - \epsilon_{m-n-1}^{(0)})}
        \propto \Delta \beta^{\frac{m}{2} - n},
\end{multline}
where $\hat{V} = f(\hat{a} + \hat{a}^\dagger)$.
This coincides with the expression $\omega_R(E) \propto \Delta e^{-S_\mathrm{tunn}(E)}$ 
where $S_\mathrm{tunn}(E)$ is defined by \eqref{tunneling_exp_2}. According to both formulas,
$\omega_R \sim \Delta \beta^{\frac{m}{2} - n}$. Moreover, not only the power--law dependence 
on $\beta$ coincides but also the numerical coefficient which is accurately derived in the
Appendix~\ref{tunneling_vs_pert_theory}.
Therefore tunneling and multi--photon transitions are just the same effects. 
Also let us note that for bistable driven systems the Keldysh parameter $\gamma_K$ logarithmically
depends on the external field amplitude $f$ whereas in the case of ionization of atoms by 
strong electromagnetic field $\gamma_K \sim f^{-1}$.

\section{The effect of degeneracy on kinetics}
The fact that the eigenstates of the quantum driven nonlinear oscillator can be superpositions
of quasiclassical states from regions $1$ and $3$ has a strong effect on kinetics described by the rate
equation. When each state can be uniquely attributed to a single region of the phase space, 
in the limit of large number of excitation quanta the rate equation transforms to
the classical Fokker--Planck equation in quasienergy representation. 
However, this is not the case when $m$ is close to an integer.
The actual eigenstates are the superpositions of the quasiclassical states from regions $1$ and $3$, 
which breaks the crucial assumption under which the Fokker--Planck equation
is derived.

However, even when $m$ is close to integer, there exists a quasiclassical limit 
of the rate equation corresponding to large $m$
which has also the form of the classical Fokker--Planck equation
in the quasienergy space. It is obtained under assumption that the occupations
$P^+_n$ and $P^-_n$ of states $|n_+\rangle$ and $|n_-\rangle$ (see Eq. \eqref{pm_states}) 
still depend smoothly on $n$. The rate equation in terms of $P^s_n$, $s=\pm$, takes form
\begin{equation}
    \label{rate_eq_pm_0}
    \begin{gathered}
        \frac{\p P_{n}^s}{\p t} = \sum_{n's'} w_{nn'}^{ss'} P_{n'}^{s'}- w_{n'n}^{s's} P_{n}^s.
    \end{gathered}
\end{equation}
where the transition probabilities $w_{nn'}^{ss'}$ between 
the states $|n_s\rangle$, $|n'_{s'}\rangle$ are defined by general formula \eqref{rate_eq}, and the
states $|n_s\rangle$ are defined by \eqref{pm_states}.
In the equation \eqref{rate_eq_pm_0}, it is possible to perform a gradient expansion 
of $P_n^\pm$.
To perform such a procedure,
it is convenient to rewrite the equation \eqref{rate_eq_pm_0} in a slightly different form:
\begin{equation}
    \label{rate_eq_pm_1}
    \begin{gathered}
        \frac{\p P_{n}^+}{\p t} = \sum_{n'} W^+_{nn'} P_{n'}^+ - W^+_{n'n}P_{n}^+ + 
                w^{+-}_{nn'} P_{n'}^- - w^{-+}_{nn'}P_{n'}^-,\\
        \frac{\p P_{n}^-}{\p t} = \sum_{n'} W^-_{nn'} P_{n'}^- - W^-_{n'n}P_{n}^- + 
                w^{-+}_{nn'} P_{n'}^+ - w^{+-}_{nn'}P_{n'}^+,\\
    \end{gathered}
\end{equation}
with the newly defined coefficients $W^+_{nn'} = w^{++}_{nn'} + w^{-+}_{nn'}$ and
$W^-_{nn'} = w^{--}_{nn'} + w^{+-}_{nn'}$. They are expressed via matrix elements between the
states from regions 1 and 3 as
\begin{widetext}
    \begin{equation}
        \label{degen_fp_coef}
        \begin{gathered}
            W^{+}_{nn'} = (N+1)\left(c_{n'}^2| a_{n_1n'_1} |^2+
                         s_{n'}^2 |a_{n_3n'_3} |^2 \right)
                        + N\left(c_{n'}^2 |a_{n'_1n_1}|^2+
                         s_{n'}^2 |a_{n'_3n_3}|^2\right),\\
            W^{-}_{nn'} =  (N+1)\left(s_{n'}^2| a_{n_1n'_1} |^2+
                         c_{n'}^2 |a_{n_3n'_3} |^2 \right)
                        + N\left(s_{n'}^2 |a_{n'_1n_1}|^2+
                         c_{n'}^2 |a_{n'_3n_3}|^2\right),\\
            w^{+-}_{nn'} = (N+1)(c_n^2s_{n'}^2 |a_{n_1n'_1}|^2 +
                          s_n^2c_{n'}^2 |a_{n_3n'_3}|^2 - 
                  2c_n c_{n'}s_n s_{n'}\Re{a_{n_1n'_1}a_{n'_3n_3}^*})\\
                           + N (c_n^2s_{n'}^2 |a_{n'_1n_1}|^2 +
                          s_n^2c_{n'}^2 |a_{n'_3n_3}|^2
                  - 2c_n c_{n'}s_n s_{n'}\Re{a^*_{n'_1n_1}a_{n_3n'_3}}),\\
            w^{-+}_{nn'} = (N+1)(s_n^2c_{n'}^2 |a_{n_1n'_1}|^2 +
                          c_n^2s_{n'}^2 |a_{n_3n'_3}|^2 - 
                  2s_n s_{n'}c_n c_{n'}\Re{a_{n_1n'_1}a_{n'_3n_3}^*})\\
                           + N (s_n^2c_{n'}^2 |a_{n'_1n_1}|^2 +
                          c_n^2s_{n'}^2 |a_{n'_3n_3}|^2
                  - 2s_n s_{n'}c_n c_{n'}\Re{a^*_{n'_1n_1}a_{n_3n'_3}}).\\
        \end{gathered}
    \end{equation}
\end{widetext}
Using the rate equations in the form \eqref{rate_eq_pm_1}, it is easy to understand the structure
of the quasiclassical limit of equations \eqref{rate_eq_pm_1}.
Following the derivation
of the Fokker--Planck equation which was in detail described in \cite{Maslova2019}, it is clear that
after gradient expansion the terms $\sum_{n'} W^{\pm}_{nn'} P_{n'}^\pm - W^{\pm}_{n'n}P_{n}^\pm$
transform to the expressions
$\frac{1}{T} \frac{\p}{\p E}\left[\vartheta K^\pm P^\pm+Q D^\pm \frac{\p P^\pm}{\p E}\right]$ 
where $K^\pm(E)$ and $D^\pm(E)$ describe the drift and the diffusion correspondingly 
in quasienergy space.
The terms $\pm(w^{+-}_{nn'} P_{n'}^- - w^{-+}_{nn'}P_{n'}^+)$
describe tunneling transitions between the states $|n_+\rangle$ and $|n_-\rangle$. 
Thus, the whole system of equations takes the form
\begin{equation}
    \label{fokker_planck_w_tunnel}
    \begin{gathered}
        \frac{\p P^{\pm}}{\p t} = \frac{1}{T}\frac{\p J_{\pm}}{\p E}+ 
            \Lambda_{\mathrm{tunn}}^\pm.\\
    \end{gathered}
\end{equation}
where
\begin{equation}
        J_\pm = \vartheta K^\pm + QD^\pm\frac{\p P^\pm}{\p E},
\end{equation}
and $\Lambda_\mathrm{tunn}^{+(-)}$ are the terms responsible for tunneling:
\begin{multline}
        \label{Lambda_definition}
        \Lambda^{+(-)}(E) \equiv 
        \int_{E_\mathrm{sep}}^{E_1} dE'T(E') \left[w^{+-(-+)}_{EE'}P^{-(+)}(E')\right.\\ 
        \left.-w^{-+(+-)}_{E'E}P^{+(-)}(E)\right].\\
\end{multline}

The system \eqref{fokker_planck_w_tunnel} 
is exactly the Fokker--Planck equation with tunneling term from 
\cite{Maslova2019}. 
The features of the equations \eqref{fokker_planck_w_tunnel} can be understood from the structure of
the eigenstates which was described in Sec.~\ref{sec:states_structure}, see Eq.~\eqref{pm_states} 
and Fig.~\ref{fig:states_structure}. As was mentioned in Sec.~\ref{sec:states_structure},
the hybridization of the quasiclassical states from different regions of the phase portrait is strong
for quasienergies below $E_c$ and it is very small for quasienergies above $E_c$. This
defines the behavior of the coefficients $K^\pm(E)$, $D^\pm(E)$ and terms
$\Lambda_\mathrm{tunn}^{\pm}(E)$
in \eqref{fokker_planck_w_tunnel} which is different for 
$E < E_c$ and $E > E_c$. For $E > E_c$
there is almost no hybridization of states from regions 1 and 3. Therefore, the
drift and diffusion coefficients 
of the Fokker--Planck equation 
$K^\pm = K_{1,3}$, $D^\pm = D_{1,3}$ and
the tunneling transition rate is small: $\Lambda^\pm_\mathrm{tunn} T \ll 1$. On the contrary, for 
$E > E_c$ the actual eigenstates are symmetric and antisymmetric superpositions of the 
quasiclassical states from different regions. Using the expressions \eqref{degen_fp_coef} and
keeping in mind that in the considered case $s_n \approx c_n \approx 1/\sqrt{2}$,
it is obvious that 
the drift and diffusion coefficients for $|n_\pm\rangle$ states are
$K^+ \approx K^- \approx \frac{1}{2}(K_1 + K_3)$, 
$D^+ \approx D^- \approx \frac{1}{2}(D_1 + D_3)$, and the tunneling rates are large and almost equal:
$\Lambda_{\mathrm{tunn}}^{+}T \approx \Lambda_\mathrm{tunn}^- T \sim 1$. 

From the considerations of the previous paragraph, it is easy to obtain the stationary distribution.
As for $E < E_c$ the drift and diffusion coefficients and tunneling transition rates
are almost equal for states $|n_\pm\rangle$ and the tunneling transition rate is large, 
the stationary probability densities $P^+$ and $P^-$ are almost
equal too, $P^+ \approx P^- \approx \bar{P}$. 
Thus, the stationary probability density $\bar{P}$ is obtained from the condition of zero 
flow $J^\pm(E)$ and is given by the expression
\begin{equation}
    \label{stat_distr_degen_below_Ec}
    \bar{P}(E) = \bigexp{-\frac{\vartheta}{Q} \int_{E_\mathrm{sep}}^E \frac{K_1 + K_3}{D_1 + D_3} dE'},
    \quad E_\mathrm{sep} < E < E_c.\\
\end{equation}
For $E_c < E < E_1$, 
the tunneling term in \eqref{fokker_planck_w_tunnel} is small, and
the stationary distributions $P^+$($P^-$) can be obtained by perturbation theory in
$\Lambda_\mathrm{tunn}^{\pm}$ as it was done in \cite{Maslova2019}. 

Let us define 
\begin{equation}
    \begin{gathered}
        P^\pm(E) = P^\pm_0(E) + \delta P^\pm(E),\\
    \end{gathered}
\end{equation}
where $P^{+(-)}_0(E)$ are stationary distributions without tunneling term and 
$\delta P^{+(-)}$ are the first--order corrections caused by tunneling terms.
Then
\begin{equation}
    \label{stat_distr_degen}
    \begin{gathered}
        P^{+}_0(E) = \bar{P}(E_c)
        \bigexp{- \frac{\vartheta}{Q} \int_{E_c}^{E} \frac{K_1}{D_1} dE'},\quad
                E_c < E < E_1\\
        P^{-}_0(E) = \bar{P}(E_c)
        \bigexp{-\frac{\vartheta}{Q} \int_{E_c}^{E} \frac{K_3}{D_3} dE'}\quad
                E_c < E < \infty.\\
    \end{gathered}
\end{equation}
The coefficients in \eqref{stat_distr_degen} are defined by the continuity condition at $E = E_c$.
The tunneling corrections have the form
\begin{equation}
    \label{stat_distr_degen_corr}
    \begin{gathered}
        \delta P^+ = -P_0^+\int_{E_\mathrm{sep}}^{E} \frac{dE'}{QD_1(E')P_0^+}
                \int_{E_1}^{E'} \Lambda_\mathrm{tunn}^+(E'') T(E'') dE'',\\
        \delta P^- = P_0^-\int_{E_\mathrm{sep}}^{E} \frac{dE'}{QD_3(E')P_0^-}
                \int_{E'}^{\infty} \Lambda_\mathrm{tunn}^-(E'') T(E'') dE'',\\
    \end{gathered}
\end{equation}
where $\Lambda_\mathrm{tunn}^\pm$ are expressed by equations \eqref{Lambda_definition} with
$P_0^{\pm}(E)$.

\begin{figure}
   \centering
   \includegraphics[width=\linewidth]{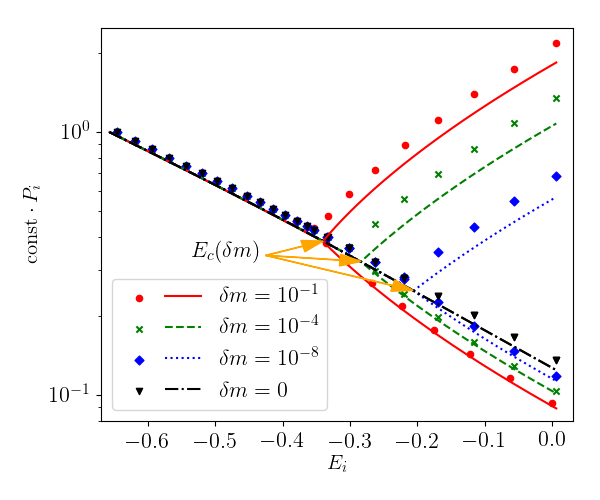}
   \caption{The distribution functions at
             $\sqrt{\beta/\beta_\mathrm{crit}} = 0.2$, $N_\mathrm{th} = 4$, 
             $m = 30 + \delta m$ for different small $\delta m$. The exact quantum 
             distributions are denoted by red circles, green crosses, blue diamonds and black 
             triangles for $\delta m = 10^{-1}, 10^{-4}, 10^{-8}, 0$. 
             The quasiclassical approximations distribution functions given by 
             Eq.~\eqref{stat_distr_degen_below_Ec} and Eq.~\eqref{stat_distr_degen} 
             are shown with red solid, green dashed, blue dotted and black dash--dotted lines 
             for different values of $\delta m$ correspondingly. The orange arrows 
             indicate the position of critical quasienergy $E_c(\delta m)$.
             }
    \label{fig:degen_distribs}
\end{figure}

On Fig.~\ref{fig:degen_distribs}, the distribution functions obtained from analytical formulas
\eqref{stat_distr_degen}, \eqref{stat_distr_degen_corr} 
are compared with those obtained by numerical solution of the equation
\eqref{rate_eq_pm_0}, \eqref{rate_eq_pm_1} 
for different values of $\delta m$. It is evident that the critical quasienergy
$E_c$ shifts towards $E_\mathrm{sep}$ with increasing $\delta m$, according to the
Eq.~\eqref{E_c_equation}. Also it can be seen that for exactly degenerate quasienergy levels
$\delta m = 0$, the system remains close to the stable state 2 which is squeezed \cite{Maslova2019}. 
In this case, the states 
corresponding to the regions 1 and 3 are equally occupied and the occupation probabilities are
exponentially small.

The ratio between the probability densities in two stable stationary states equals
\begin{multline}
    \label{p2_p1_ratio}
    \frac{Q}{\vartheta}\ln{\frac{P_2(E_2)}{P_1(E_1)}} = 
                    \int_{E_2}^{E_\mathrm{sep}} \frac{K_2(E')}{D_2(E')} dE'\\
                    +\int_{E_\mathrm{sep}}^{E_c} \frac{K_1(E') + K_3(E')}{D_1(E')+D_3(E')} dE'\\
                    +\int_{E_c}^{E_1} \frac{K_1(E')}{D_1(E')} dE' +
                    \ln{\left[1 + \frac{\delta P^+}{P_0^+}\right]}.
\end{multline}
The tunneling correction for $E > E_c$ doesn't lead to any qualitative effects because it is of 
order $\alpha/\Delta$ comparing to $P_0^\pm$. 

\begin{figure}
    \centering
    \includegraphics[width=\linewidth]{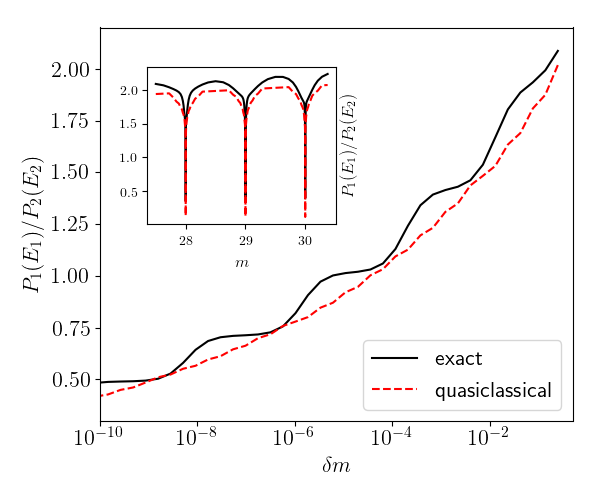}
    \caption{The ratio of probability densities in the stationary states $1$ and $2$ at
             $\sqrt{\beta/\beta_\mathrm{crit}} = 0.2$, $N_\mathrm{th} = 4$, 
             $m = 30 + \delta m$. The exact value obtained from transition matrix
             diagonalization for quantum oscillator is compared with analytical formula
             \eqref{p2_p1_ratio}. 
             On the inset, the dependence of probabilities ratio 
             $P_2(E_2)/P_1(E_1)$ on $m$ is shown in linear scale for 
             the same parameters. At integer values of $m$, there are exponentially 
             narrow dips.}
    \label{fig:p2_p1_ratio}
\end{figure}

On Fig.~\ref{fig:p2_p1_ratio}, the analytical formula
\eqref{p2_p1_ratio} is compared with the numerical result for the dependence of 
$P_1(E_1)/P_2(E_2)$ on $\delta m$. The analytical formula fits the 
numerical result quite well.  However, the discreteness of the
quasienergy levels manifests itself in the smooth steps in the dependence 
$P_1(E_1)/P_2(E_2)$ on $\delta m$ which are not reproduced by \eqref{p2_p1_ratio}.
These steps can be explained by the fact that the crossover energy $\epsilon_c$ can take only 
discrete values. Thus, for 
\begin{equation}
    e^{-S_\mathrm{tunn}(E_{n})} < \delta m < e^{-S_\mathrm{tunn}(E_{n+1})}
\end{equation}
the effective position of $E_c$ remains the same. When 
$\delta m \approx e^{-S_\mathrm{tunn}(E_{n+1})}$, 
the value of $E_c$ abruptly changes from 
$E_n$ to $E_{n+1}$. This explains the presence of steps on
Fig.~\ref{fig:p2_p1_ratio}. 
The width of the steps in the logarithmic scale is 
$S_\mathrm{tunn}(E_{n}) - S_\mathrm{tunn}(E_{n+1})
\approx 2\pi T_\mathrm{im}(E_n)/T(E_n)$. In the latter expression, we recognize the previously 
defined Keldysh parameter $\gamma_K$. 
For $n$ much smaller than $m$ and $\beta \ll \beta_\mathrm{crit}$, 
$\gamma_K \sim \ln{\frac{1}{\beta}}$. 

On the inset in Fig.~\ref{fig:p2_p1_ratio}, the dependence of probabilities ratio 
$P_1(E_1)/P_2(E_2)$ on $m$ is shown in linear scale for $\sqrt{\beta/\beta_\mathrm{crit}} = 0.2$.
For non--degenerate case, when $m$ is far from an integer, the state with smaller amplitude and 
quasienergy $E_1$ is the most probable. However, when $m$ becomes close to an integer, the 
occupation of the state with quasienergy $E_1$ abruptly drops, and the value at the minima
is exponentially small for large $m$ (see the dips on the inset of Fig.~\ref{fig:p2_p1_ratio}). 
The width of the dips in linear scale is also exponentially small.

\section{Conclusions}
We considered the non--equilibrium statistics and kinetics of the model of resonantly 
driven quantum nonlinear oscillator interacting with dissipative environment. We found out that
the non--equilibrium statistics and kinetics are strongly modified when quasienergy states are
nearly degenerate
which occurs at integer or half--integer detuning--nonlinearity ratio. 
In particular, the occupation of the classical stable state 
with smaller amplitude is strongly reduced. So, in the case of exactly degenerate quasienergy levels
the system always occupies the state with higher amplitude which is squeezed.

The coefficients of the Fokker--Planck equation which describes the kinetics in the quasiclassical 
limit are very sensitive to the structure of 
eigenstates of the system's Hamiltonian. 
We found out that in the case of integer or half--integer detuning--nonlinearity ratio, 
which corresponds to the exact multi--photon resonance between the genuine energy levels of the 
unperturbed nonlinear oscillator, the quasienergy states from different regions of the phase
space simultaneously hybridize and form symmetric and antisymmetric superpositions. This fact
can be proven by applying a special symmetry transformation to the Hamiltonian.
Also we revealed that when the quasienergy levels of the system are nearly degenerate, 
a new important critical quasienergy parameter $\epsilon_c$ emerges. 
Below $\epsilon_c$, all quasienergy states are superpositions 
of the quasiclassical states from regions 1 and 3, and above $\epsilon_c$, 
the quasienergy states correspond to either region 1 or 3.  We found out that 
the coefficients of the Fokker--Planck equation which describes the quasiclassical 
kinetics of the oscillator in almost--degenerate case have different behavior above and below
$\epsilon_c$. In particular, tunneling term is large below $\epsilon_c$ and exponentially small 
above $\epsilon_c$. Also the drift and diffusion coefficients are not affected by tunneling 
above $\epsilon_c$ whereas below $\epsilon_c$ they are strongly modified.  
The distribution functions and the ratio between occupations of the classical stable states 
calculated analytically fit well the numerical results.

We generalized Keldysh theory for ionization of atoms in electromagnetic field for bistable systems.
It was demonstrated that Keldysh parameter defined as the ratio of <<tunneling time>> to the
quasiclassical period of motion along the phase trajectory is large in the bistability region for
external field intensity smaller than the critical value. So the multi--photon transition and
tunneling between different regions of the phase space can be treated as the same effects. This fact
was proved by direct calculation of transition amplitude 
using both tunneling and perturbation theory
approach. Also we revealed that the
Keldysh parameter for the considered system depends logarithmically on the amplitude $f$ 
of the external field. On the contrast, in the case of multi--photon ionization of 
atoms the Keldysh parameter is inversely proportional to the amplitude of the external field.

\begin{acknowledgements}
    This work was supported by RFBR grants 19--02--000--87a and 18--29--20032mk and by a grant of
    the Foundation for the Advancement of Theoretical Physics and Mathematics ''Basis''.
\end{acknowledgements}

\bibliographystyle{apsrev4-1}
\input{enhanced_excitation_190805.bbl}

\appendix

\section{The coefficients of the classical Fokker--Planck equation}
\label{appendix:quasiclassical_quantization}

The coefficients of the classical FPE 
are defined as line integrals along the classical trajectories of the nonlinear oscillator:
\begin{equation}
    \label{fp_coefficients}
    \begin{gathered}
        K_{i}(E) = \frac{i}{2}\oint a\,da^* - a^*\,da,\\
        D_{i}(E) = 
            \frac{i}{2}\oint \frac{\p H}{\p a} da - \frac{\p H}{\p a^*} da^*,\\
        T_{i}(E) = \int da^*da\,  \delta(E - H(a^*, a)).\\
    \end{gathered}
\end{equation}
The classical trajectory is a contour line of the classical Hamiltonian
\begin{equation}
    \label{class_ham}
    H_\mathrm{cl} = -\Delta |a|^2 + \frac{\alpha}{2}|a|^2 + f(a + a^*),
\end{equation}
where $a = \frac{q + ip}{\sqrt{2}}$, and $q, p$ is a pair of canonically conjugate variables.
The index $i$ denotes the region of the phase space according to Fig.~\ref{fig:phase_portrait}.
The coefficient $K_i(E)$ is proportional to the adiabatic invariant of the trajectory defined as
\begin{equation}
    \label{ad_inv_1}
    n_i(E, f) = \frac{1}{2\pi}K_i(E) = \frac{1}{2\pi}\oint p\,dq.
\end{equation}

\begin{figure}[h]
    \includegraphics[width=0.9\linewidth]{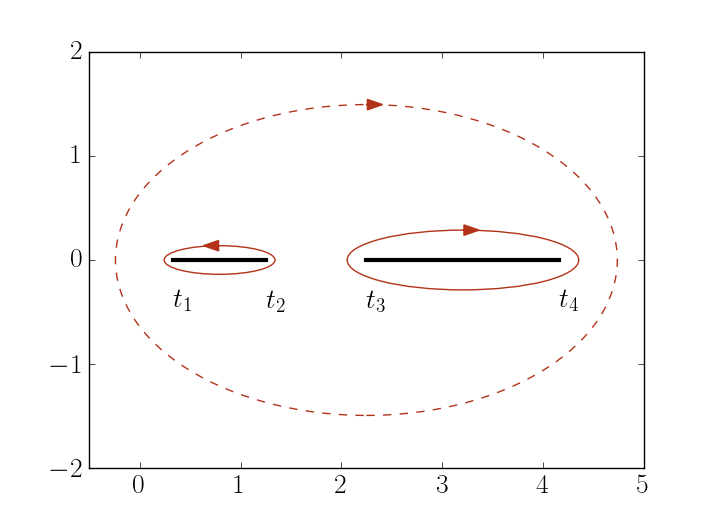}
    \caption{The contours of integration in \eqref{ad_inv_integral} corresponding to 
             $n_1(\epsilon, f)$ (left contour) and
             $n_3(\epsilon, f)$ (right contour) are shown on the complex plane of $t$. The difference
             $n_3 - n_1$ is expressed as an integral over outer contour (dashed line)}
    \label{fig:contours}
\end{figure}

The expressions \eqref{fp_coefficients} and \eqref{ad_inv} can be rewritten as two--dimensional
integrals in $p, q$ plane with a Dirac delta function 
as in expression for $T_i$. Then, it is convenient to
use variables $q, t$ instead of $q, p$ where $t = q^2 + p^2$. Then the coefficients are transformed
to one--dimensional integrals by $t$. 
Now let us focus on the expression for adiabatic invariant:
\begin{equation}
    \label{ad_inv_integral}
    n_i(E, f) = \frac{m}{2}
            \oint_{C_i} \frac{dt}{4\pi}\frac{3t^2/16 - t/4 +E/2}
            {\sqrt{2f^2t-\left(E+\frac{t}{2}-\frac{t^2}{8}\right)^2}}.
\end{equation}
Using this expression, we will prove the identity of Eq.~\eqref{ad_inv_identity}.
The contour of integration in \eqref{ad_inv_integral} 
depends on the region of the phase space in which the 
trajectory lies. In the range of energies corresponding to the region 2 of the phase space, 
the polynomial has only two real roots, and the contour of integration encloses them. 
In the range of energies corresponding to the region $1$, there are 4 real roots:
$t_1 < t_2 < t_3 < t_4$ (see Fig.~\ref{fig:contours}). The range $t_1 < t < t_2$ ($t_3 < t < t_4$) 
corresponds to the trajectories from the region 1 (3). Thus, the contour of integration for
$n_1$ ($n_3$) encloses $t_1$ and $t_2$ ($t_3$ and $t_4$).

By deformation of the contour, it is easy to show that 
$n_3(\epsilon, f) - n_1(\epsilon, f)$ is expressed
through residue of integrand \eqref{ad_inv_integral} on infinity. Expanding the integrand
of \eqref{ad_inv_integral} in $t^{-1}$, one gets the desired identity
\begin{equation}
    n_3(\epsilon, f) - n_1(\epsilon, f) = m = \frac{2\Delta}{\alpha}.
\end{equation}

\section{The proof of the symmetry of the perturbative corrections}
\label{appendix:symmetry}
Here we prove that the perturbation theory corrections in $f$ 
to the quasienergies $\epsilon_n^{(0)}$ of 
the eigenstate with $n$ excitation quanta of t
he model \eqref{quant_bist_ham} at $f = 0$ are symmetric with 
respect to replacement $n \to m-n$. 
This fact was mentioned in \cite{Risken1988} and \cite{Dykman2005}, but 
the authors didn't give any proof to this fact.

For several low--order corrections, this can be verified 
by straightforward calculation, as for the second--order correction
\eqref{2_ord_corr}:
\begin{equation}
     \epsilon^{(2)}_n = \frac{f^2}{\alpha}\cdot  \frac{(m + 1)}{(m - 2n)^2  - 1}.
\end{equation}
However, it is necessary to clarify what the expression $\epsilon_{m-n}^{(k)}$ means for 
non--integer $m$ because the corrections $\epsilon_n^{(k)}$ are defined only for integer 
$n$ which has the meaning of the number of excitation
quanta. Thus, for non--integer $m$ the identity 
$\epsilon_n^{(k)} = \epsilon_{m-n}^{(k)}$ holds 
only for formal expressions. Up to now, we didn't give any meaning to $\epsilon_\nu^{(k)}$ for 
non--integer $\nu$ except as analytic continuation of perturbation theory formulas.

However, it is in fact possible to give direct meaning to $\epsilon_\nu$ and $\epsilon_\nu^{(k)}$
at non--integer $\nu$. For 
that, we should formally assume that the Hamiltonian \eqref{quant_bist_ham} acts on the space
of all possible real <<numbers of excitation quanta>> $\nu$ with operators $a$, $a^\dagger$ defined as
follows:
\begin{equation}
    \label{a_generalization}
    \begin{gathered}
        \langle\nu|a|\nu+1\rangle = \langle\nu+1|a^\dagger|\nu\rangle = \sqrt{\nu}.\\
    \end{gathered}
\end{equation}
Then, for $f = 0$ each state $|\nu\rangle$ is an eigenstate with energy 
$\epsilon_\nu^{(0)} = \nu(m-\nu)$.
For $f \ne 0$, they become coupled with 
$|\nu\pm1\rangle, |\nu\pm2\rangle, \dots,|\nu\pm k\rangle\dots$. If none of the states
$|\nu\pm k\rangle$ are degenerate with $\nu$ (equivalently, $2\nu - m$ is non--integer), the
amplitudes of $|\nu\pm k\rangle$ remain small at small $f$, and it is possible to define
continuous $f$--dependent energy $\epsilon_\nu(f)$. This is the energy of the eigenstate 
which evolves from $|\nu\rangle$ after adiabatic switching of the perturbation. For 
integer $2\nu - m$, the energy $\epsilon_\nu(f)$ can't be defined that way 
because of degeneracy between $|\nu\rangle$ and $|m-\nu\rangle$.

The series of perturbation theory for $\epsilon_\nu$ in the cases of integer and non--integer $\nu$
are completely identical because of definition \eqref{a_generalization}. Thus, the 
claim that $\epsilon^{(k)}_n = \epsilon^{(k)}_{m-n}$ follows from even more general statement 
$\epsilon_\nu = \epsilon_{m-\nu}$.

We prove the identity $\epsilon_\nu = \epsilon_{m-\nu}$ in several steps. First, it is obvious from the 
previous considerations that $\epsilon_\nu(f)$ is an eigenvalue of the operator 
\begin{equation}
    \label{gen_ham}
    \mathcal{H}_\nu = \frac{\alpha}{2}\cdot\sum_{\sigma - \nu \in \mathbb{Z}} 
        \sigma(\sigma-m) |\sigma\rangle\langle \sigma| + 
        f\sqrt{\sigma}(|\sigma-1\rangle\langle \sigma| + |\sigma\rangle\langle \sigma-1|)
\end{equation}
which corresponds to the state $|\nu\rangle$. Analogously, $\epsilon_{m-\nu}(f)$ arises from 
the operator $\mathcal{H}_{m-\nu}$. 
For convenience in the later discussion, we change the numeration of basis vectors 
in $\mathcal{H}_{m-\nu}$ so that $|\sigma\rangle$ becomes $|m - \sigma\rangle$.
After such relabeling,
\begin{multline}
    \mathcal{H}_{m-\nu} = \frac{\alpha}{2}\sum_{\sigma - \nu \in \mathbb{Z}} 
        \sigma(\sigma-m) |\sigma\rangle\langle \sigma| + \\
        f\sum \sqrt{m-\sigma}(|\sigma+1\rangle\langle \sigma|+|\sigma\rangle\langle \sigma+1|).
\end{multline}
Both $\mathcal{H}_\nu$ and $\mathcal{H}_{m-\nu}$ act on a 
single space with a set of basis vectors $|\sigma\rangle$ with such $\sigma$ that
$\sigma - \nu$ is integer. We should emphasize that 
$\mathcal{H}_\nu$ and $\mathcal{H}_{m-\nu}$ are substantially 
different and could not be transformed to each other by any permutation of eigenvectors.

However, there exists a nontrivial linear operator $T$ which transforms $\mathcal{H}_\nu$ to 
$\mathcal{H}_{m-\nu}$:
\begin{equation}
    \label{transform_1}
    \mathcal{H}_\nu = \mathcal{T}\mathcal{H}_{m-\nu} \mathcal{T}^{-1}.
\end{equation}
It has the form
\begin{equation}
    \label{transform_2}
    \mathcal{T} = UTU'^{-1},\\
\end{equation}
where
\begin{equation}
    \label{transform_3}
    \begin{gathered}
        U  = \sum_{\sigma}\sqrt{\Gamma(\sigma+1)}|\sigma\rangle\langle\sigma|\\
        U' = \sum_{\sigma}\sqrt{\Gamma(m-\sigma)}|\sigma\rangle\langle\sigma|\\
        T = \bigexp{\frac{2f}{\alpha}\sum |\sigma\rangle\langle\sigma+1|}
    \end{gathered}
\end{equation}
The identities \eqref{transform_1}, \eqref{transform_2}, \eqref{transform_3} are checked
by direct calculation. 

The existence of the operator $\mathcal{T}$ is possible only because of special form
of $\epsilon_\nu = \frac{\alpha}{2}\nu(m-\nu)$. 
For any other dependence of $\epsilon_\nu$ on $\nu$, no analogous
operator can be found this way.
So, the symmetry property expressed by $\mathcal{T}$ is a special feature of 
Kerr--like nonlinearity.

The equivalence of Hamiltonians $\mathcal{H}_\nu$ and $\mathcal{H}_{m-\nu}$ proves that
the energies $\epsilon_\nu(f)$ and $\epsilon_{m-\nu}(f)$ are equal when $2\nu - m$ is not integer. However,
we are interested in the case of integer $m$ and integer numbers of excitation quanta. For 
this case, one should utilize degenerate perturbation theory to find the energies.
Nevertheless, the corrections to energies of $|n\rangle$ and $|m-n\rangle$ obtained by 
degenerate perturbation theory are just the same as in non--degenerate perturbation
theory up to the order $f^{|m-2n|}$. This happens  
because the leading contribution to composite matrix element (multi--photon Rabi frequency)
between $|n\rangle$ and $|m-n\rangle$ is a product of $|m-2n|$ matrix elements of the 
perturbation $\hat{V}$ (see \eqref{pert_splitting}, \eqref{pert_splitting_app}).
For series of non--degenerate perturbation theory the identity for $k$--th order corrections
$\epsilon_{n}^{(k)} = \epsilon_{m-n}^{(k)}$ holds even for integer $m$ and $n$, if $k < 2|m-2n|$. For 
$k \ge 2|m-2n|$, the corrections of non--degenerate perturbation theory don't make sense 
because of a zero in denominator, which is a manifestation of degeneracy. This means that
the degeneracy of $|n\rangle$ and $|m-n\rangle$ is lifted only in the order $|m-2n|$, and 
the energy splitting happens only due to multi--photon Rabi oscillations:
$\Delta \epsilon_{n,m-n} = 2\omega_R^{n,m-n} + o(f^{|m-2n|})$. 

\section{The identity of tunneling splitting and multi--photon transition amplitude}
\label{tunneling_vs_pert_theory}
The multi--photon Rabi splitting between the quasienergy states $|n\rangle$ and $|m-n\rangle$ is
given by formula
\begin{multline}
    \label{pert_splitting_app}
    \omega_R^{n,m-n} = \frac{V_{n,n+1}\dots V_{m-n-1,m-n}}
                            {(\epsilon_n^{(0)} - \epsilon_{n-1}^{(0)})
                            \dots(\epsilon_n^{(0)} - \epsilon_{m-n-1}^{(0)})} =\\
        =\alpha \left(\frac{2f}{\alpha}\right)^{m-2n}
            \!\!\!\!\!\!\!\!\!\!
            \frac{1}{(m-2n-1)!^2}\sqrt{\frac{(m-n)!}{n!}}
\end{multline}
At large values of $n$, $m$, it is possible to approximate the factorials using the Stirling formula.
Then one gets the following expression for $\omega_R^{n,m-n}$:
\begin{multline}
    \ln{\frac{\omega_R^{n,m-n}}{\Delta}} = \frac{m}{2}\left[(1-r)\ln{\frac{1}{\beta}}
           + (1-r)(2\ln{2}-3) \right.\\ 
         \left. + 4(1-r)\ln(1-r) -\frac{1}{2}((2-r)\ln(2-r) - r\ln{r})\right], 
\end{multline}
where $r = 2n/m$. In this form, it is easy to compare it with tunneling splitting given by
Eq.~\eqref{quasiclassical_splitting} and Eq.~\eqref{tunneling_exp}. 
At small $\beta$, the tunneling action \eqref{tunneling_exp} can be approximated as 
\begin{equation}
    \label{s_tunn_small_beta}
    \begin{gathered}
        S_\mathrm{tunn} = \frac{m}{2}\left[\sqrt{1 + 2E}\ln{\frac{2}{\beta}} + 
            \int_{x_1}^{x_2} \ln{\frac{E + \frac{x^2}{2} - \frac{x^4}{8}}{x}}xdx\right]\\
        x_{1,2} = \sqrt{2 \mp 2\sqrt{1 + 2E}}
    \end{gathered}
\end{equation}
For small external force the quasienergy $E$ is related with number of excitation quanta by formula
$E = \frac{2n}{m}(\frac{n}{m} - 1) = r^2/2 - r$. Evaluating the integral in \eqref{s_tunn_small_beta}
and substituting the expression for $E$ via $r$, it is easy to obtain that in current approximations
$S_\mathrm{tunn} = \ln{(\omega_R^{n,m-n}/\Delta)}$. Thus, at $\beta \ll \beta_\mathrm{crit}$ 
perturbation theory
is consistent with tunneling approach.
\end{document}

%% file: enhanced_excitation_190805.bbl
%